\documentclass[twocolumn]{aastex62}
\usepackage[nointegrals]{wasysym} 
\usepackage{threeparttable}
\usepackage{tabularx}
\usepackage{xcolor}
\usepackage{amsmath}
\usepackage{lineno}
\usepackage{stix}

\graphicspath{{./}{figures/}}

\newcommand\eaps{Department of Earth, Atmospheric, and Planetary Sciences, Massachusetts Institute of Technology, 77 Massachusetts Avenue, Cambridge, MA 02139, USA}
\newcommand\mki{Department of Physics and Kavli Institute for Astrophysics and Space Research, Massachusetts Institute of Technology, Cambridge, MA 02139, USA}
\newcommand\aeroastro{Department of Aeronautics and Astronautics, Massachusetts Institute of Technology, Cambridge, MA 02139, USA}

\submitjournal{ApJ}

\shorttitle{Light to shadow}
\shortauthors{Shapiro, Seager, Solanki et al.}
\newcommand\mps{Max-Planck-Institut f\"ur
Sonnensystemforschung, Justus-von-Liebig-Weg 3, 37077 G\"ottingen, Germany}
\begin{document}

\title{The Curious Case of Dark Faculae on M Dwarf Stars}

\author[0000-0002-8842-5403]{Shapiro, A.I.}
\affiliation{Institute of Physics, University of Graz, Graz, Austria}
\correspondingauthor{Alexander~I.~Shapiro}
\affiliation{\mps}
\email{alexander.shapiro@uni-graz.at}

\author[0000-0002-6892-6948]{Sara Seager}
\affiliation{\eaps}
\affiliation{\mki}
\affil{\aeroastro}
\correspondingauthor{Sara Seager}
\email{seager@mit.edu}

\author[0000-0002-3418-8449]{Sami~K.~Solanki}
\affil{\mps}
\affil{School of Space Research, Kyung Hee University, Yongin, Gyeonggi 17104, Republic of Korea}

\author[0000-0002-6087-3271]{Nadiia~Kostogryz}
\affil{\mps}

\author[0000-0002-0929-1612]{Veronika~Witzke}
\affiliation{Institute of Physics, University of Graz, Graz, Austria}

\author[0000-0002-3243-1230] {Krishnamurthy~Sowmya}
\affiliation{Institute of Physics, University of Graz, Graz, Austria}

\author[0000-0002-6568-6942]{Tanayveer~Bhatia}
\affil{\mps}

\author[0000-0001-8217-6998]{Yvonne~C.~Unruh}
\affiliation{Department of Physics, Imperial College London, London SW7 2AZ, UK}

\author[0000-0002-6568-6942]{Robert~Cameron}
\affil{\mps}

\begin{abstract}
The study of exoplanet atmospheres has brought a renewed interest in stellar astrophysics. Specifically, stellar magnetic activity from the host star contaminates exoplanet atmosphere transmission spectra, as evidenced by observations from the James Webb Space Telescope. The stellar surface magnetic features in the form of dark spots caused by large concentrations of surface magnetic fields and more diffuse and extended faculae caused by small-scale magnetic field concentrations change the apparent size of the star in a wavelength-dependent way. This spot and faculae contribution to the change in transit depth contaminates the planetary signal. Here we study the transition from bright faculae on G- and K-dwarfs to dark faculae on M-dwarfs.
This dark appearance of faculae is in significant contrast to the conventional picture that faculae are brighter than the quiet star region as they are on the Sun. We use the 3D radiative magnetohydrodynamics code MURaM to simulate faculae, and calculate the faculae spectra with the MPS-ATLAS radiative transfer code. We present a qualitative explanation for the transition from dark to bright faculae attributing it to shallower flux tubes and reduced vertical temperature gradients at the surfaces of M dwarfs relative to the Sun. 
\end{abstract}


\section{Introduction}\label{sect:Intro} 
Small-scale magnetic fields are known to affect the surface brightness of the Sun and are also believed to influence the surface brightness of other cool main-sequence stars with convective zones. The stellar regions covered by such fields are called faculae, but are also sometimes referred to as plage (which are the chromospheric counterparts of photospheric faculae) or bright spots. On the Sun, facular regions are brighter than surrounding quiet regions at most wavelengths and disk positions.  Solar facular regions are often found in the vicinity of dark sunspots, the latter being concentrations of larger amounts of magnetic flux,  but faculae can also appear independently of spots.  Faculae occupy larger areas on the solar surface (up to about 5\%) compared to spots (up to about 1\%) and are the main driver of solar irradiance variability on the timescale of the Sun's 11-year activity cycle \citep[see, e.g. review by][]{ARAA2013}.

Faculae and spots are also believed to be present on the surfaces of other cool stars. A number of phenomena (such as stellar photometric and spectral variability) caused by stellar magnetic features have been actively studied for decades \citep[see, e.g., a detailed summary in][and references therein]{Gibor2019}. 
The recent renewed interest in the magnetic features of cool stars is largely driven by their suspected impact on the interpretation of exoplanetary data. In particular, the advent of transmission spectroscopy with the James Webb Space Telescope has further intensified efforts to understand the brightness contrasts of stellar magnetic features. Magnetic features that remain unocculted during a transit offset the light curve \citep[see recent review by][]{Rackham2023} and effectively alter the apparent stellar radius \citep[see discussion in][]{Seager2024}, thereby contaminating the transmission spectra of exoplanets.



The key information needed for removing stellar contamination is the spectra of spots and faculae. While some observational constraints are available for stellar spots \citep[see, e.g.,][]{sveta2005}, there have been no direct observations of faculae on stars other than the Sun. In the absence of a better approach, the exoplanetary community has broadly adopted the solar paradigm for representing faculae, even on stars that are very different from the Sun. Namely, faculae are assumed to be brighter than the quiet (non-magnetic) stellar regions, and their spectra have been routinely approximated by the spectra of hotter stars (computed based on 1D stellar models). 

In this paper, we identify the limitations of the solar faculae paradigm and show that it breaks down for a broad class of stars, most notably, red dwarfs. The physical mechanisms that make solar faculae appear bright are complex and rather specific to the Sun and stars not too different from it. In the solar atmosphere, magnetic fields are concentrated in small, roughly vertical flux tubes with strengths of 1–2 kG, located in the intergranular lanes between convective granules \citep{Solanki1993}. These magnetic concentrations are strong enough to inhibit convective energy transport \citep[][]{Rempel2011}, thereby suppressing convective heating—similar to what occurs in sunspots.
Solar faculae appear bright primarily because magnetic pressure depresses the optical surface within flux tubes into deeper layers of the near-surface solar atmosphere \citep[][]{Spruit1976, DePontieu2006, ARAA2013}. Because the temperature increases rapidly with depth in the Sun, 
so-called hot walls surround the facular flux tubes. When faculae are observed from the side (e.g., close to the limb), the bounding hot walls are usually directly visible through the optically thin interiors of the flux tubes, leading to a significant increase in facular brightness. Furthermore, the radiation from the hot walls heats the interiors of facular flux tubes.

The observed solar facular brightness compared to the quiet star regions is actually a combination of several factors: the magnitude of the flux tubes' depression into the surface; {the vertical temperature gradient at and just below the stellar surface} (which defines the temperature of hot walls); the ability of the hot walls to heat the interior of the flux tubes (which, in turn, depends on the temperature of the hot walls, radii of the flux tubes, and the opacity of the gas within the flux tubes). On the Sun, the combination of all these factors leads to  bright faculae, despite the inhibition of convective heating inside the flux tubes. 

All in all, the many factors that generate and control the visibility of solar faculae  are intertwined with the properties of the host star, which implies that flux tube properties may differ on other stars.   In other words, we should not assume the bright appearance of solar faculae can simply be directly transferred to other stars. In the absence of direct observations the best way to account for effects associated with the visibility  of faculae is to rely on comprehensive 3D radiative magnetohydrodynamics (MHD) simulations. These simulations have recently revealed that facular spectra are very different from those obtained by 1D models \citep{Witzke2022}. This is not surprising since hot walls are intrinsically a 3D effect.  Quite surprisingly,  these simulations have indicated that faculae might get dark towards cooler stars \citep{Beeck2015, Salhab2018, Norris2023}.

In this paper we use the 3D radiative MHD code MURaM  \citep[][]{Vogler2005} and the radiative transfer code MPS-ATLAS \citep{Witzke2021} to calculate spectra of faculae for G2, K2, M0, and M4 main sequence stars. We show that faculae are dark on M4 stars compared to the quiet star regions and provide a qualitative explanation for the transition from bright to dark faculae.


\begin{figure*}[ht]
\includegraphics[scale=0.6]{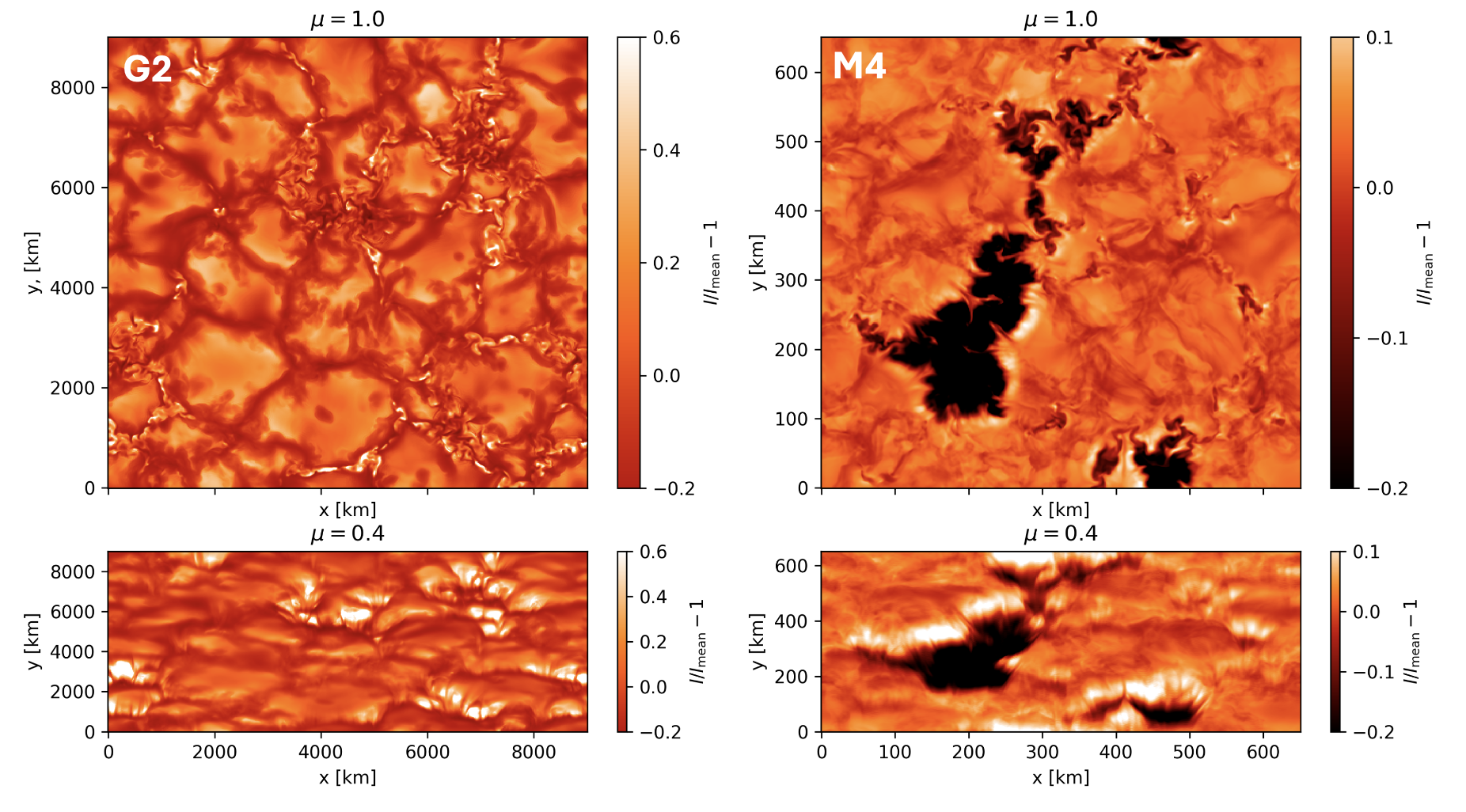}
\caption{Simulated faculae on G2 and M4 dwarf stars at 600 nm observed at disk center ($\mu=1$, top panels) and an intermediate disk position ($\mu=0.4$, bottom panels). The images are normalized to the averaged intensity from the entire image. Faculae structures are computed with MURaM followed by intensity calculations using MPS-Atlas.  The magnetic field in faculae on the M4-dwarf (right panels) leads to the formation of  dark regions (reminiscent of solar pores). In contrast, the magnetic field on the G2 dwarf (left panels) causes the formation of bright points in the intergranular lanes.}
\label{fig:images}
\end{figure*}

\section{Results}\label{sect:result}
The main finding of this work is that faculae are darker than the quiet (i.e. non-magnetic) stellar regions on M4 dwarf stars while on M0 dwarfs they are similar in brightness to the quiet stellar regions. The appearance and structure of faculae differ significantly between M4 dwarfs and G2 dwarfs (like the Sun). On M4 dwarfs, faculae consist of dark, pore-like structures, whereas on G2 dwarfs, faculae appear as bright points or sheet-like structures (Figure~\ref{fig:images}) in striking agreement with solar observations performed with high spatial resolution \citep[][]{Lites2004,Keller2004,Carlsson2004, Kuridze2025}. The dark appearance of the faculae on M4 dwarfs stands in sharp contrast to the conventional assumption, commonly adopted in studies of stellar contamination, that faculae are brighter than the quiet photosphere irrespective of stellar spectral class. In this section, we assess the implications of this result for the contamination of transmission spectra.

\subsection{Facular Contamination on Transmission Spectra}\label{subsect:result_TS}

To place our main result in the context of exoplanet atmosphere studies, we present illustrative calculations of facular contamination of exoplanet transmission spectra $\epsilon (\lambda)$. These calculations are performed using the standart approach \citep[][]{Rackham2018, Rackham2023, Seager2024} 
\begin{equation}\label{eq:def} 
   \epsilon (\lambda) \equiv \frac{R_{\rm cont}(\lambda)^2 - R_{\rm true}(\lambda)^2}{R_{\rm true}(\lambda)^2} = \frac{F_{\rm quiet}(\lambda) - F_{\rm active}(\lambda)}{F_{\rm quiet}(\lambda)}. 
\end{equation}
Here,  $R_{\rm true}(\lambda)$ is the true radius of the exoplanet, which would be retrieved if the planet were orbiting a quiet (non-magnetic) star. $R_{\rm cont}(\lambda)$ represents the contaminated radius of the exoplanet, i.e., the radius retrieved from the transmission spectra of a planet orbiting an active star (a star with magnetic features). $F_{\rm active}(\lambda)$ and $F_{\rm quiet}(\lambda)$ are the spectral fluxes from active and quiet stars, respectively.   {We note that Eq.~\ref{eq:def} is derived assuming that the transit chord does not cross active regions \citep[see, e.g.,][for a detailed discussion and the derivation of Eq.~\ref{eq:def}]{Rackham2018, Rackham2023, Seager2024}. This simplification is well suited for our purposes, as it allows us to directly connect the facular contrast to the contamination $\epsilon(\lambda)$.}

\begin{figure*}[ht]
\includegraphics[scale=0.2]{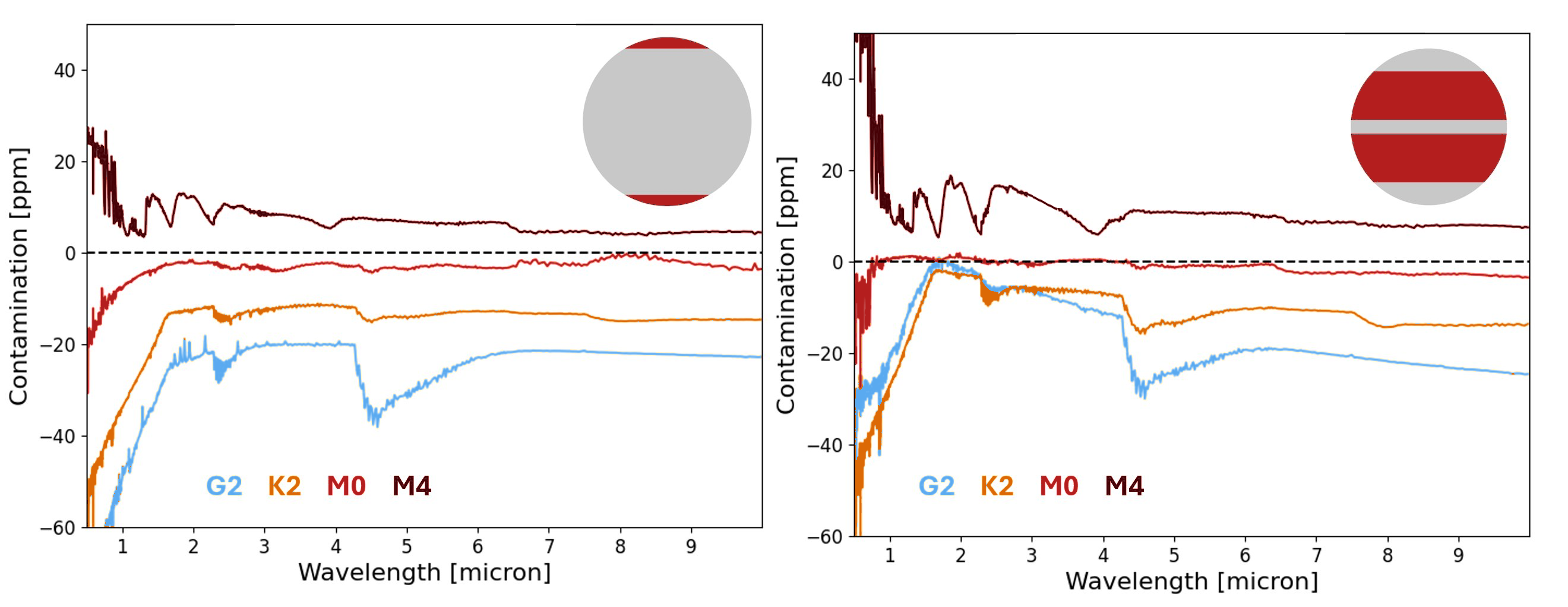}
\caption{The contamination of transit spectra by faculae. Shown are offsets (defined as the difference between ``observed''  and ``true'' values of the transit depth) introduced by faculae  on G2, K2, M0, and M4 dwarf stars. Calculations are performed for a 10,000 ppm planet transit and two facular distributions: ``polar'' (left panel,  faculae are populated into two polar caps (north and south) with latitude $>60 ^{\circ}$) and ``equatorial' (right panel,  faculae are populated into two bands between latitudes $\pm 5 ^{\circ}$ and $\pm 40 ^{\circ}$. In each case, only a fraction of the polar caps and equatorial belts is covered by faculae, chosen such that the total stellar surface coverage is 20\%. Independently of the assumed surface distribution of faculae the contamination becomes positive for M4 dwarfs (i.e. faculae get dark).}
\label{fig:cont}
\end{figure*}

If all systematic factors \citep[e.g., stellar limb darkening, see, e.g.,][]{Rosa2024} are properly accounted for, the dependence of the exoplanet radius, $R_{\rm true}$, on wavelength, $\lambda$, would reveal the presence of an exoplanet atmosphere. However, the wavelength-dependent contamination by the faculae can mimic or obscure this dependence, posing a significant challenge for interpreting transmission spectra.

\begin{figure*}[ht]
\includegraphics[scale=0.45]{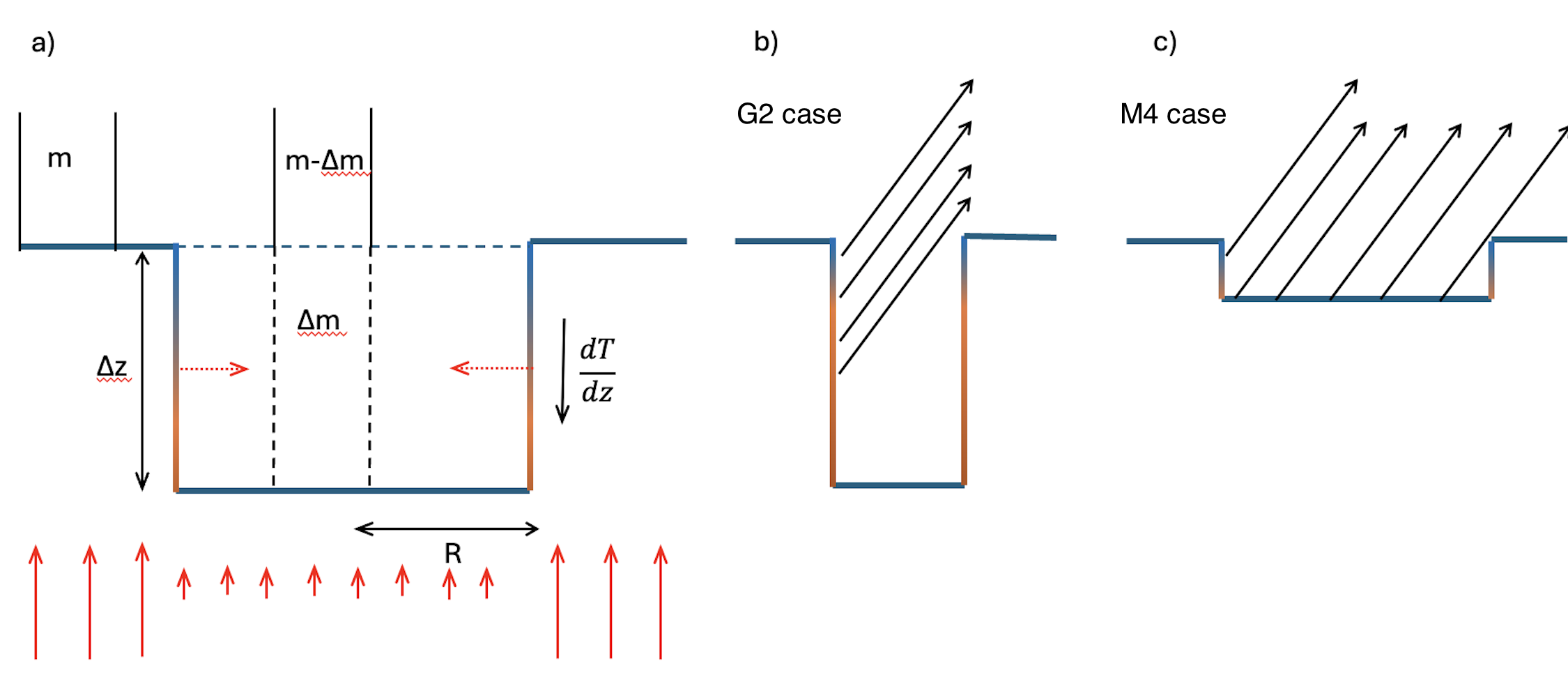}
\caption{Panel a: Schematic representation of the main features of magnetic flux tube affecting its visibility and brightness.  The flux tube is defined by depth $\Delta z$, radius $R$, column mass $m$. The temperature gradient is $dT/dz$.  The depth $\Delta z$ is to first order defined by the reduction of the column mass at the surface, $\Delta {\rm m}$, due to the magnetic pressure ({see Sect.~\ref{subsect:geometry}}). Red arrows represent convective energy flux.  Panel b: G2 stars' deep flux tubes with hot walls are visible from most of the  lines of sight {(represented by black arrows)}. Panel c: M4 dwarf star shallow flux tubes have walls that are less visible than in the case shown in panel b (black arrows).}
\label{fig:scheme}
\end{figure*}

Our calculations reveal that, contrary to the current paradigm, faculae on the surfaces of M4 dwarf stars increase the observed planetary radius (positive contamination, i.e. $\epsilon (\lambda)>0$), resulting in absorption features in transmission spectra. M0-dwarfs represent a transition from positive contamination on cooler stars to negative contamination ($\epsilon (\lambda)<0$) in warmer stars. 

This result is illustrated in Fig.~\ref{fig:cont}, where we present facular contamination for a central transit with a magnitude of 10,000 ppm. Since the brightness of faculae strongly depends on their position on the stellar disk, we present contamination curves for two limiting cases of facular distribution on the stellar disk  \cite[taking these distributions from][]{Shapiro2014}. While the surface distribution  slightly affects the spectral profile of the contamination (compare the left and right panels of Fig.~\ref{fig:cont}), it does not alter our main result: faculae are dark compared to the quiet stellar regions on M4 dwarfs so that they cause an increase in the exoplanetary radius derived from observations. {The amplitude of the contamination strongly increases toward shorter wavelengths reaching about 100 ppm for M4 dwarfs and 400 ppm for G2 dwarfs at 300 nm (not shown on the plot which stops at 600 nm to focus on the wavelength domain observed by the JWST).}

\subsection{Stellar Model Fluxes}\label{subsect:model}
The contamination curves in Fig.~\ref{fig:cont} have been obtained using the following expressions for the spectral fluxes of quiet and active stars, respectively:
\begin{equation}\label{eq:quiet}
 F_{\rm quiet}(\lambda) = \int_{\rm disk}   I_Q (\lambda, \vec{r} ) \, {\rm d} \Omega,
\end{equation}
\begin{equation}\label{eq:active}
 F_{\rm active}(\lambda) = F_{\rm quiet}(\lambda) +  \int_{\rm active} \left ( I_{\rm fac}  (\lambda, \vec{r} )- I_Q (\lambda, \vec{r} ) \right ) \, {\rm d} \Omega.
\end{equation}
The integration in Eq.~\ref{eq:quiet} is performed over the visible stellar disk. $I_Q (\lambda, \vec{r} )$ is the spectral intensity from the quiet stellar surface along the direction $\vec{r} $ and ${\rm d} \Omega$ is the differential of the solid angle around the direction $\vec{r}$. Similarly, $I_{\rm fac}(\lambda, \vec{r} )$ is the spectral intensity from faculae. The integration in Eq.~\ref{eq:active} is performed over the parts of the stellar disk covered by faculae, e.g. the stellar polar regions for the contamination curves shown in the left panel of Fig.~\ref{fig:cont} and the near-equatorial parts for the contamination curves shown in the right panel of Fig.~\ref{fig:cont}. 

We use 3D MHD  simulations of faculae and quiet stellar regions with the MURaM code and compute the emergent spectra, $I_Q (\lambda, \vec{r} )$ and $I_{\rm fac}(\lambda, \vec{r})$ needed for Eqs.~\ref{eq:quiet}--\ref{eq:active} with the MPS-ATLAS code. The calculations with the MPS-ATLAS code are performed in the ray-by-ray approach, i.e. by interpolating the density, temperature, and pressure of the 3D MURaM simulation box into parallel rays and solving radiative transfer equation along these rays. This approach allows for accurate accounting of the 3D structure of the quiet Sun and, most importantly, the faculae, whose 3D structure plays a critical role in determining their visibility. We refer to the recent studies of \cite{Norris2023, Kostogryz2024, Witzke2024, Smitha2024} for a more detailed discussion of the ray-by-ray approach and its applications.

The MURaM simulations used in this study are taken from multiple sources but follow a homogeneous setup. Thus, the simulations of the quiet stellar regions include the action of a small-scale turbulent dynamo, which is currently thought to result in the most realistic simulations of quiet-star regions \citep[see, e.g.,][for a number of observational tests]{Kostogryz2024, Witzke2024}.  Faculae are simulated by adding an initially uniform vertical magnetic field to the MURaM simulation box representing quiet regions and letting the simulations evolve and stabilize \citep[see the detailed description in][]{Beeck2015, Witzke2022, Kostogryz2024}. We note that the exact value of the added field has little effect on the properties of the individual flux tubes mainly affecting the number of flux tubes in the simulation box \citep{Beeck2015}. For convenience we adopt the value of 200 G for our calculations.

For G2-dwarfs we took simulations of quiet stellar regions from \cite{Witzke2023}. We took facular G2-dwarf simulations from \cite{Witzke2022} using their 200 G setup.  The simulations of the quiet regions on K2- and M0-dwarfs have been taken from \cite{bhatia2022}. The faculae on K2- and M0-dwarfs have been newly simulated by adding 200 G flux to the \cite{bhatia2022} quiet-star simulations. The simulations of quiet and facular regions on M4-dwarfs have been performed following the setup of  \cite{bhatia2022} and \cite{Witzke2022}.

\section{Qualitative Explanation of Dark Faculae on M Dwarf Stars}\label{sect:explanation}
The combination of MURaM simulations of flux tubes with subsequent ray-by-ray calculations of emergent spectra using the MPS-ATLAS code, as presented in Sect.~\ref{sect:result}, accounts for the complex physics of the interaction between matter, magnetic fields, and radiation required to describe the visibility of faculae. In particular, MURaM simulations self-consistently account for the depression of the stellar surface to deeper layers in facular flux tubes, inhibition of convection by the magnetic field, as well as the depression of faculae into deeper stellar layers, leading to the formation of hot walls and the radiative heating of flux tubes. The ray-by-ray spectral synthesis with the MPS-ATLAS code, in turn, accounts for the intricate effects introduced by flux tubes on the emergent spectra, such as the complex temperature and density behavior within the flux tube (both horizontally and vertically) and the visibility of the hot walls and cool bottom of the flux tube as a function of wavelength and viewing angle. The combination of these effects determines the dependence of the facular spectrum on wavelength and disk position for stars with different effective temperatures, resulting in the contamination profiles showcased in Fig.~\ref{fig:cont}.

Although physically accurate, the numerical simulations alone do not provide a clear understanding of the main physical concepts that lead to the significantly different appearance of faculae on G and M dwarfs. To address this, we complement our numerical simulations with a qualitative description to pinpoint and explain the main physical effects causing faculae to appear darker than quiet-star regions on M dwarfs but brighter on G  and K dwarfs.

\subsection{Two Competing Effects Define the Visibility of Faculae}\label{subsect:mechanisms}
Stellar magnetic features, i.e. faculae and spots, are caused by the magnetic field protruding through the stellar surface in the form of complex, three dimensional  structures. While MURaM simulations capture these structures, to describe them conceptually we rely here on a  simplified picture, that of flux tubes \citep{Solanki1993}:  regions on the stellar surface that contain strong (1--2 kG at the surface) predominantely vertical magnetic fields \citep[see Fig. 5 and detailed explanations in ][]{ARAA2013}.  The flux tubes pass through the stellar surface directed roughly vertically and display a variety of horizontal sizes. On the Sun they can be as narrow as 10--100 km  and as wide as several tens of thousands of km (for comparison the typical size of solar granules is about 1000 km), with smaller flux tubes emerging much more often than larger ones \citep[see, e.g.][]{Thornton2011, Smitha2017, Natasha2021}, and hence being also much more common at the solar surface \citep{Parnell2009, Anusha2017}. 

The vertical magnetic field in the flux tubes leads to two important effects that define the visibility of the faculae. First, the vertical magnetic field stops the horizontal gas flows in the stellar photosphere. This is because charged particles cannot move perpendicular to the magnetic field and neutral particles are, in turn, collisionally connected to the charged particles. 
The horizontal flows are an essential part of the convective pattern (specifically, of the overturning part of the convective cells). As a result, the convective flux of energy into the flux tube from below is significantly reduced \citep{Rempel2011}. In the absence of other effects, flux tubes would, therefore, be cooler than the surrounding non-magnetic regions, resulting in dark faculae.

However, a second effect comes into play and may alter this outcome. The magnetic field causes the optical surface within the flux tube to be depressed into deeper and, consequently, hotter regions of the star. This occurs because the total (i.e. thermal and magnetic pressure) pressure values within and outside the flux tube must be equal at the same horizontal layer. The magnetic pressure adds to the total pressure in the flux tube, lowering the gas pressure and, consequently, the gas opacity. As a result the optical surface in the flux tubes corresponds to deeper stellar layers than non-magnetic quiet-star regions (see Figure~\ref{fig:scheme}). {This phenomenon, first identified by \citet{Wilson1774} and subsequently quantified by \citet{Gokhale1972} (see also the monograph by \citealt{BrayLoughhead1964}), is known as the Wilson depression, and its magnitude in the solar case can reach up to 600 km \citep{Loptein2018}.}

The Wilson depression, in turn, affects the visibility of faculae through two mechanisms. First, it leads to a 'geometric brightening effect' when flux tubes are seen from the side: deeper (and thus hotter) layers hidden on the non-magnetic star become visible (see Figure~\ref{fig:scheme}b), manifesting themselves as ‘hot walls’ surrounding the flux tube. This mechanism is well-known in solar physics \citep[see, e.g. the review by][]{Solanki1993} and it is well reproduced by our solar MURaM simulations \citep[][]{Norris2017, Witzke2022}. It also operates in our simulations of M4 dwarfs, as evident in the bottom right panel of Figure~\ref{fig:images}, where hot, bright walls are clearly visible along the northern boundaries of dark, pore-like structures.  
Interestingly, hot walls are also visible even when flux tubes are viewed directly from above (i.e., at the disk center). They manifest as brightening near the edges of pore-like structures (see the upper right panel of Figure~\ref{fig:images}). This effect is not captured by our simplified schematic in Figure~\ref{fig:scheme} and arises from the expansion and corrugation of the hot walls. 
A second mechanism associated with bright walls is the radiative heating of flux tube interiors, whereby radiation from the hot walls penetrates and warms the flux tube’s interior. All in all, both mechanisms driven by the hot walls act to enhance facular brightness.

On the Sun, faculae appear brighter than the surrounding non-magnetic regions at most wavelengths and disk positions. This indicates that the combined action of two ‘brightening’ mechanisms caused by the Wilson depression generally overcompensates for the ‘darkening’ effect of convection inhibition in horizontally small flux tubes that form faculae. Both ‘brightening’ effects are important. For example, in the ultraviolet spectral domain and in the cores of spectral lines, solar faculae appear brighter than the surrounding quiet regions—even when observed at disk center, where the ‘geometric brightening effect’ effect of direct visibility of hot walls is expected to be minimal. This implies that radiative heating from the hot walls overcompensates for the inhibition of convection at the depths where the radiation is formed.  At the same time, when observed in the visible and infrared continuum, solar faculae appear dark at the disk center \citep[e.g.][]{Foukal1990, Wang1998} and only become bright near the limb, owing to the “geometric brightening effect” effect of hot walls \citep[a phenomenon well reproduced by the MURaM simulations, see, e.g.,][]{Norris2017}.

The existence of bright faculae on the Sun was first explained in the classic work of \cite{Spruit1976} 
who performed a sophisticated semi-analytical assessment of the flux tubes' visibility. Accurately extending this to stars with non-solar fundamental parameters would be quite cumbersome, considering the very different atmospheric structures, levels of convection, depression of the stellar surface in magnetic features, and atmospheric opacities. With today's comprehensive 3D MHD and radiative transfer codes (such as the MURaM and MPS-ATLAS codes used in this study), the community has moved away from physical semi-analytic descriptions. We still, however, benefit from a conceptual description to qualitatively illustrate the transition from bright faculae on the Sun to dark faculae on M-dwarfs. We emphasize that this is done here for purely illustrative purposes and cannot replace the simulations presented in Sect.~\ref{sect:result} and elsewhere.

\subsection{Dependence of Facular Brightening Efficiency on Stellar Temperature and Metallicity}\label{subsect:result_TFE}
Instead of following the detailed derivation in \cite{Spruit1976} we will approximate the brightening effect from hot walls via the dimensionless proxy ${\cal HW}$ (short for ${\cal H}$ot ${\cal W}$all):

\begin{equation}
  \label{eq:proxy_der_1}
   {\cal HW} = \left (\frac{\alpha}{1-\alpha}  \right )^2 \cdot \left ( \frac{dT}{dP} \frac{P}{T} \right ) \bigg|_{\rm surface}.
\end{equation}
A higher value of ${\cal HW}$ indicates stronger hot-wall effects and, consequently, brighter faculae. The factor $\alpha$ in the first term of Equation~\ref{eq:proxy_der_1} quantifies the efficiency of the {\it convective collapse} (see below) and is formally defined as the ratio between the magnetic and total  pressure within the flux tube, $\alpha \equiv {P_{\rm mag}}/{P_{\rm tot}}$. The second term on the right-hand side of Equation~\ref{eq:proxy_der_1} is well known in stellar atmosphere theory \citep[see, e.g., Chapter 8 of][]{Mihalas1978} and is often referred to as the dimensionless temperature gradient ($ {d \ln T} / {d \ln P} $). In Equation~\ref{eq:proxy_der_1}, it is evaluated at the horizontal layer corresponding to the optical surface outside the flux tube (commonly referred to in the literature as the surface corresponding to a Rosseland optical depth of unity). The temperature $T = T(z)$ and pressure $P = P(z)$ in this term are functions of height in the non-magnetic stellar atmosphere. A detailed derivation of Equation~\ref{eq:proxy_der_1} is presented in Sect.~\ref{sect:der}; here, we focus on its evaluation and on understanding its dependence on stellar parameters.

A strength of Equation~\ref{eq:proxy_der_1} is that it separates magnetic and non-magnetic effects. Magnetic effects are solely contained within the first factor, through the efficiency of convective collapse, $\alpha$. The second factor is purely a result of radiative-convective equilibrium, which determines the vertical structure of the quiet stellar atmosphere and is unaffected by magnetic field.


In atmospheres where energy is transported by radiation, the dimensionless temperature gradient can be expressed as a function of the Rosseland optical depth, $\tau_{\text{Ross}}$:
\begin{equation}\label{eq:grad}
 \frac{dT}{dP} \frac{P}{T} = \frac{1}{4} \, m_c (\tau_{\text{Ross}}) \cdot \kappa_{\text{Ross}} \cdot \frac{1}{\tau_{\text{Ross}} + 2/3},
\end{equation}
where  $m_c (\tau_{\text{Ross}}) $ is the column mass (i.e. the amount of mass per unit area above a given layer in the atmosphere) and $\kappa_{\text{Ross}}$ is the Rosseland mean opacity (per unit mass), both evaluated at the layer corresponding to the optical depth $\tau_{\text{Ross}}$. The derivation of the Equation~\ref{eq:grad} is given in Appendix~\ref{app:A}.

Equation~\ref{eq:grad} provides a reasonable estimate of the temperature gradient at the stellar surface (i.e., at the $\tau_{\rm Ross} = 1$ layer), where radiative transfer of energy generally plays a more important role than convection. It shows that the gradient at the surface  is proportional to the product of the Rosseland opacity and the column mass at the surface. This product is typically of order unity, except in cases where the opacity varies strongly with height in the atmosphere. We remind here that the opacity in Equation~\ref{eq:grad} is defined per unit mass and, thus, it is not {\it directly} influenced by the steep increase in density toward deeper layers. To illustrate why product of the Rosseland opacity and the column mass at the surface is generally near unity, consider the limiting case in which the Rosseland opacity is constant with height. In this case, the optical depth is simply the product of opacity and column mass, implying that at the $\tau_{\rm Ross} = 1$ surface, $m_c = 1 / \kappa_{\text{Ross}}$.  Consequently, the temperature gradient at the stellar surface is not expected to vary strongly with stellar parameters. The dependence of ${\cal HW}$ on stellar parameters therefore primarily reflects changes in the efficiency of convective collapse, described by the parameter $\alpha$.  

Historically, the phenomenon of convective collapse was introduced in the seminal work of \citet{Parker1978} to explain a surprising solar observation: most of the magnetic flux on the solar surface, excluding that associated with sunspots, was found to be concentrated in small flux tubes containing nearly vertical magnetic fields with remarkably uniform strengths of about 1.5 kG as first shown by \cite{stenflo1973}. Interestingly, various radiative MHD simulations have since demonstrated {that the 1.5 kG value at the surface of flux tubes is surprisingly universal: magnetic field strengths found in the flux tubes of stars spanning a broad range of fundamental parameters are very close to the solar value of 1.5 kG \citep{Beeck2015, Salhab2018, Tanay2024}. }


The gas pressure at the stellar surface strongly depends on fundamental stellar parameters (see Figure~\ref{fig:app} in the Appendix). Therefore, the universality of magnetic field strength (and thus magnetic pressure) in flux tubes implies that star-to-star variations in surface gas pressure must be compensated by corresponding variations in the efficiency of convective collapse. For instance, the surface gas pressure increases toward cooler stars (largely due to the fact that they have higher surface gravitational acceleration, $\log g$, values). This implies that the efficiency of convective collapse decreases with decreasing effective temperature. Indeed, such a decline was predicted  by \citet{Rajaguru2002}, who solved the linear eigenvalue problem formulated by \citet{Spruit1979}.

In Appendix~\ref{app:B}, we provide a brief overview of the convective collapse phenomenon along with qualitative arguments that support the idea of a universal magnetic field strength and magnetic pressure in flux tubes. Here, we compute $\alpha$ in Equation~\ref{eq:proxy_der_1} by assuming a magnetic field strength of 1.5 kG to evaluate the magnetic pressure. We use the surface gas pressure obtained from the library of non-magnetic 1D stellar atmosphere models by \citet{Kostogryz2023}.

\begin{figure}
\includegraphics[width=\linewidth]{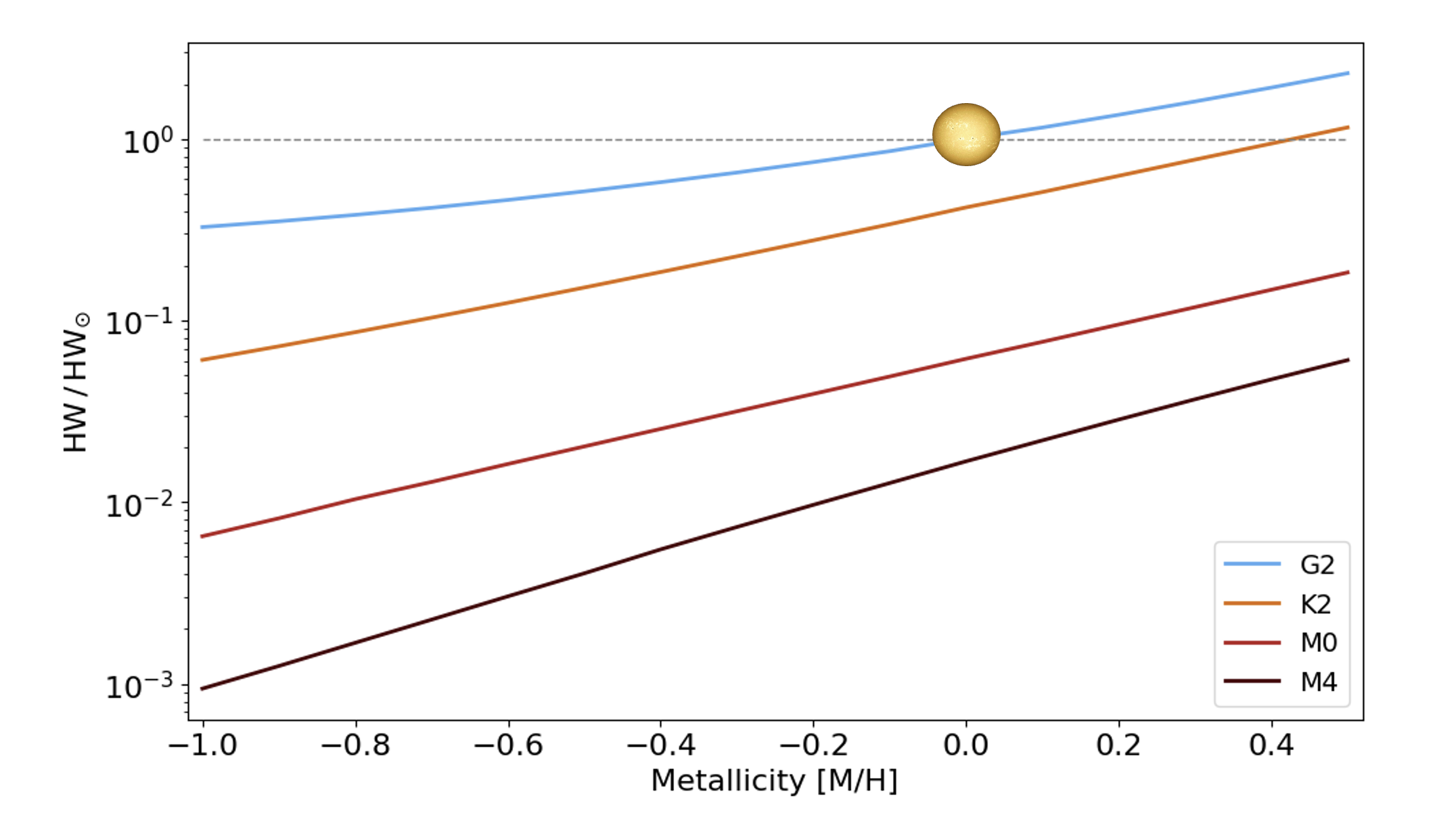}
\caption{The dependence of the ${\cal HW}$ proxy, as defined by Eq.~\ref{eq:proxy_der_1}, on stellar effective temperature and metallicity. Plotted are ${\cal HW}$ values, normalized to the solar value (${\cal HW}_{\odot}$), for four different effective temperatures and a broad range of metallicities. The  ${\cal HW}$ values decrease for both cooler stars and stars with lower metallicity, which accounts for the transition from bright to dark faculae.}
\label{fig:result}
\end{figure}

The key result of this section is that the value of ${\cal HW}$ decreases towards cooler stars (see  Fig.~\ref{fig:result}). This explains the transition from bright to dark faculae as well as the results of  Sect.~\ref{sect:result} and previous studies based on 3D radiative MHD simulations \citep[][]{Salhab2018, Norris2023}. 

Secondly, the  ${\cal HW}$ value exhibits a strong  metallicity dependence,  decreasing towards metal-poor stars, similarly as it does for M-dwarfs with solar metallicity. The phenomenon of dark faculae for metal-poor stars and general dependence of facular contrast on stellar metallicity was predicted by  \cite{Witzke2018}. It agrees with observed trends in stellar photometric variability \citep[][]{Karoff2018, Witzke2020} and recent 3D MHD simulations with the MURaM code (Witzke, priv. comm.).

The decrease of ${\cal HW}$ toward lower-metallicity stars is attributed to the reduced opacity in their atmospheres which leads to higher column mass and pressure at the surface (as more gas is required to reach $\tau_{\rm Ross} = 1$). Notably, this perspective provides a unified explanation for the appearance of dark faculae in both low-metallicity stars and cool M dwarfs: in both cases, the higher (relative to the Sun) surface pressure results in shallower flux tubes (see below) and a diminished contribution from hot walls. We expect, however, that the dependence of facular contrast on metallicity is stronger than suggested by the proxy ${\cal HW}$. This is because facular contrast is strongly amplified in spectral lines \citep[][]{Shapiro_lines, Yeo_fac1, Norris2023}. Since the contribution of spectral lines diminishes with decreasing metallicity, the facular contrast is reduced beyond what is captured by the proxy ${\cal HW}$. 

In summary, the simplified framework offered by the ${\cal HW}$ proxy provides a qualitative explanation for facular visibility, circumventing the complexity of full radiative 3D MHD simulations. While our derivation relies on 1D non-magnetic stellar models to estimate surface pressure, these models are readily available, extensively documented in the literature, and enable straightforward replication of results—unlike more complex 3D models. For instance, the dependencies shown in Figure~\ref{fig:result} were generated using only a few lines of Python code.

\section{Derivation of the Expression for the ${\cal HW}$ Proxy}\label{sect:der}
In this section, we explain why the ${\cal HW}$ proxy defined in Equation~\ref{eq:proxy_der_1} provides a means to quantify the brightening effect of hot walls. {We first show that the strength of the brightening effect from the hot walls can be quantified as:
\begin{equation}
    \label{eq:proxy}
 {brightening}  \sim \left ( \frac{\Delta z}{R} \right ) \cdot \left  (\frac{dT}{dz} \frac{\Delta z}{T_{\rm eff}} \right ), 
\end{equation}}
where $\Delta z$ is the magnitude of the Wilson depression (i.e., the depth at which the optical surface in the flux tube is depressed into the star), $R$ is the radius of the flux tube (see Figure~\ref{fig:scheme}a). $dT / dz$ is the vertical temperature gradient at the quiet stellar surface and $T_{\rm eff}$ is the effective temperature of the star. We define the coordinate system such that \( z \) increases toward deeper layers of the atmosphere so that $dT / dz$ is positive (our derivation focuses on the photosphere and subsurface layers, where the temperature increases with depth).

The first term in Equation~\ref{eq:proxy} characterizes the geometry of the flux tubes, indicating whether they are deep or shallow, while the second term captures the brightness contrast of the hot walls relative to the quiet stellar surface. We examine both terms in detail below and demonstrate that ${\cal HW}$ serves as a proxy for the two primary “brightening” mechanisms: the direct visibility of hot walls (Sect.~\ref{subsect:illusion}) and the radiative heating of the flux tube interior (Sect.~\ref{subsect:heating}). In Sect.~\ref{subsect:geometry}, we further show how Equation~\ref{eq:proxy} can be simplified to the form given in Equation~\ref{eq:proxy_der_1}.

\subsection{The direct visibility of hot walls ('geometric brightening effect') }\label{subsect:illusion}
We approximate the visibility of the hot walls by the first term in Equation~\ref{eq:proxy}, i.e., by the ratio between the magnitude of the Wilson depression and the radius  of the flux tube: $\Delta z /R$. A large value of the ratio corresponds to a deep flux tube with a small radius. In this case hot walls are easily observable even for near-vertical rays (see Figure~\ref{fig:scheme}b). In other words, hot walls can be directly seen even if faculae are located far from the stellar limb. However, such flux tubes disappear close to the limb, as the hot walls are hidden by the intervening quiet Sun. Conversely, a small value of the $\Delta z /R$ ratio corresponds to shallow flux tubes with large radii, where the flux tube interior dominates (see Figure~\ref{fig:scheme}c). For such cases, hot walls can only be glimpsed when faculae are located very close to the stellar limb. 

In addition to the purely geometrical factor, the relative brightness of the hot walls is affected by the temperature difference between the hot walls and the non-magnetic stellar surface. This difference is given by the vertical temperature gradient at the quiet stellar surface, $dT / dz$. For example, regardless of the $\Delta z /R$ ratio, there will be no hot walls in an isothermal atmosphere. A large temperature gradient would lead to bright hot walls that can affect the visibility of even a shallow flux tube. 

On average the temperature difference between hot walls and quiet stellar surface, $\Delta T$, is proportional to $dT / dz \cdot \Delta z$ (for example, it will be $dT / dz \cdot \Delta z/2$ at the depth halfway between the quiet stellar surface and the surface inside the flux tube neglecting the change of the temperature gradient with height). Approximating the brightness of hot walls and quiet stellar surface by the Stefan–Boltzmann law we can connect the relative excess of hot walls' brightness with respect to the quiet surface as $\Delta T /T$. (Because the total brightness is proportional to $T^4$ and the excess of brightness is proportional to $T^3 \cdot \Delta T$). 
All in all, $dT / dz \cdot \Delta z$ defines the temperature excess of the hot walls relative to the quiet surface.  Dividing it by $T_{\rm eff}$ gives the brightness excess of hot walls, resulting in the second term in Equation~\ref{eq:proxy}. We will use the effective temperature of a star, $T_{\rm eff}$, to get the general trends. 



\subsection{Heating of the flux tube interior by hot walls}\label{subsect:heating}
The amount of radiative energy penetrating the flux tube from the hot walls per unit time is proportional to the area of the hot walls, which is, in turn, proportional to $\Delta z \cdot R$. The energy transferred per unit time and per unit area of the hot walls is proportional to $T_{\rm hot , walls}^4 - T_{\rm flux , tube}^4 \sim T_{\rm flux , tube}^3 \cdot (T_{\rm hot , walls} - T_{\rm flux , tube})$. We can then characterize the radiative heating from hot walls, ${\cal RH}$, by the expression:
\begin{equation}\label{RT}
    {\cal RH} \sim  (\Delta z \, R ) \cdot (T_{\rm eff}^3 \, \Delta z \, \frac{dT}{dz}  )
\end{equation}
Here we approximate $T_{\rm hot \, walls}$ by $T_{\rm eff}$ and $T_{\rm hot \, walls}- T_{\rm flux \, tube}$ by the temperature excess of flux tubes introduced in Sect.~\ref{subsect:illusion}: $dT/dz \cdot \Delta z$.  Strictly speaking, this approximation holds only for optically thin flux tubes where the average gas temperature equals the star’s effective temperature. Thus, Equation~\ref{RT} ignores a rather sophisticated distribution of temperature within the flux tube and  an accurate treatment of the radiative transfer. Nonetheless, Equation~\ref{RT} captures two key effects: radiative heating from the hot walls is proportional to their area (first term in Equation~\ref{RT}) and their brightness (second term in Equation~\ref{RT}). We emphasize that our MURaM simulations do not rely on the simplifications made in the derivation of Equation~\ref{RT}. Instead, the simulations accurately account for the flux tube structure and radiative transfer.

The total energy given by Equation~\ref{RT} is emitted from the flux tube through its base (with an area proportional to $R^2$). The corresponding energy emitted from a quiet stellar region of the same surface area is proportional to $R^2 \cdot T_{\rm eff}^4$. Dividing Equation~\ref{RT} by this expression once again yields the proxy given by Equation~\ref{eq:proxy} that now quantifies  the energy flowing into and heating the flux tube from the hot walls, relative to the energy emitted by the quiet stellar surface.

\subsection{The Hydrostatic Equilibrium of 
Flux Tubes} \label{subsect:geometry}
The condition of hydrostatic equilibrium implies that the pressure at the optical surface is  $m_c g$, where $g$ is the surface gravity, and $m_c$ is the column mass at the optical surface of a star.  We note that strictly speaking the column mass and pressure are slightly varying along the optical surface (e.g. due to the convection) but this is a second order effect for our purposes,  so that we will neglect it. 

Outside the flux tube the pressure is mainly attributed to the gas thermal pressure. Within the flux tube also the magnetic pressure plays an important role. The ratio between the magnetic and total gas pressure can be expressed via the efficiency of the convective collapse, $\alpha$:
\begin{equation}
\label{eff}
    \alpha = \frac{P_{\rm mag}}{P_{\rm tot}} = \frac{B^2}{8 \pi m_c g},
\end{equation}
where $B$ is the magnetic field within the flux tube (at the optical surface). If we assume horizontal pressure balance, the value of $\alpha$ also defines the decrease of the gas pressure within the flux tube (which is reduced by a factor of $1-\alpha$). Neglecting the temperature difference between gas within and outside of the flux tube we can then directly connect the decrease of density to the the decrease of pressure:
\begin{equation}
    \rho'=(1-\alpha) \rho,
\end{equation}
where $\rho'$ and $\rho$ are densities within and outside of the flux tube, respectively.


Let us now write the equation for the equality of the pressures  outside of and within the flux tube at the geometrical height corresponding to the optical surface ($z=0$ surface, see Figure \ref{fig:scheme}a). Outside of the flux tube:
\begin{equation}
\label{ballance}
m_c^{\rm out} g = m_c^{\rm within}  g + \frac{B^2}{8 \pi} \equiv m_c^{\rm within}  g  + \alpha \,   m_c^{\rm out}   g.
\end{equation}
Here, $m_c^{\rm out}$ and $m_c^{\rm within}$  are column masses at $z=0$ outside of and inside the flux tube, respectively.  Equation \ref{ballance} allows connecting the drop of the column mass within the flux tube, $ \Delta m_c \equiv m_c^{\rm out} - m_c^{\rm within} $, to the efficiency of the convective collapse: $\Delta m_c = \alpha \, m_c^{\rm out} $. For example, in the case of a $\alpha=1$ a full evacuation of the gas from the flux tube occurs and $\Delta m_c = m_c^{\rm out}$. 




We can now write the approximate expression for the magnitude of the Wilson depression as:
\begin{equation}
    \Delta z = \frac{\Delta m_c}{\rho'} = \frac{m_c}{\rho} \cdot \frac{\alpha}{1-\alpha} \equiv h \cdot \frac{\alpha}{1-\alpha},
\end{equation}
where $h$ is the photospheric scale height. 

{Thus, we can now rewrite Eq.~(\ref{eq:proxy}) in the following way:
\begin{equation}
 \label{eq:proxy_der_2}
 {brightening}  \sim  \left (     \frac{h}{R} \cdot \frac{\alpha}{1-\alpha} \right) \cdot \left  (\frac{dT}{dz}   \cdot \frac{\alpha}{1-\alpha}     \cdot   \frac{h}{T_{\rm eff}} \right ). 
\end{equation}}

To further simplify the equations we assume that the mean radius of the flux tubes on a star scales with the average horizontal size of granules on this star (we note that this assumption is introduced here for the sake of simplicity and the  deviations from it lead to an extra darkening of faculae on M-dwarf, see below). The radius of granules, in turns, scales with the photospheric scale height, $h$ {so that our assumption boils down to $h \sim R $}.

{By putting this into Eq.~(\ref{eq:proxy_der_2}) and recalling that $h \sim T_{\rm eff}/(g \mu)$ (with $g$ the stellar surface gravity and $\mu$ the mean molecular mass), $dz = dP/(g \rho)$ (with $\rho$ the gas density), and the ideal gas equation in the form $\rho/\mu \sim P/T$, we can rewrite Eq.~(\ref{eq:proxy_der_2}) as 
\begin{equation}
    \label{eq:proxy_der_next}
 {brightening} \sim \left (\frac{\alpha}{1-\alpha}  \right )^2 \cdot \left ( \frac{dT}{dP} \frac{P}{T} \right )\bigg|_{\rm surface} \equiv {\cal HW} .
\end{equation}
Here we also neglected the small ($\approx 6\%$) difference between $T_{\rm eff}$ and the surface temperature $T_{\rm surf}$, using the Eddington approximation $T_{\rm surf} \approx 1.06\, T_{\rm eff}$. This concludes the derivation of Eq.~(\ref{eq:proxy}).  }

We note that the assumption that the radius of a flux tube scales with the horizontal size of granules does not capture the pronounced morphological differences in faculae between MURaM simulations of G2 and M4 stars. On G2 stars, faculae consist of multiple small flux tubes located in intergranular lanes, whereas on M4 stars, they are concentrated in a few large flux tubes (see Figure~\ref{fig:images}) that push granules away. 
The exact origin of this morphological change remains to be investigated \citep[it is also less pronounced in the structure of the magnetic field, see][]{Tanay2026}. In any case, the relative increase in flux tube size compared to granule size on M4 stars further contributes to the darkening of faculae.

\section{Conclusions}\label{sect:Concl}
The facular regions on cool M-dwarfs appear darker than the surrounding quiet, non-magnetic stellar surface. This behavior stands in stark contrast to the solar case, where faculae are brighter than the quiet Sun across nearly all wavelengths and disk positions.

Our findings are based on comprehensive 3D radiative magnetohydrodynamic simulations of faculae using the MURaM code, combined with spectral synthesis performed with the MPS-ATLAS code. In addition, we carried out a simplified, qualitative analysis grounded in basic physical principles. This approach reveals that the somewhat counterintuitive dark appearance of faculae on cool M-dwarfs (and metal-poor dwarfs) can be explained by the higher surface gas pressure in these stars compared to the Sun. This leads to flatter, less corrugated surfaces, which suppress the brightening  associated with hot-wall effects.

These results imply that bright surface features observed on cool M-dwarfs may arise from mechanisms unrelated to faculae. We will explore this possibility in more detail in a forthcoming study.

\vspace{5mm}

\appendix

\section{Dimensionless temperature gradient}\label{app:A}
Here we calculate the dimensionless temperature gradient that appears in Equation~\ref{eq:proxy_der_1}:
\begin{equation}
\label{def}
\nabla \equiv \frac{d \ln T}{d \ln P} = \frac{dT}{dP} \cdot \frac{P}{T}. 
\end{equation}
Let us write the equation of the hydrostatic equilibrium:
\begin{equation}\label{eq:HE}
dP =  g \cdot \rho \cdot dz,
\end{equation}
and the ideal gas law:
\begin{equation}\label{eq:ideal}
P = \frac{\rho {\cal R} T}{\mu}.
\end{equation}
Eqs.~\ref{eq:HE}--\ref{eq:ideal} allow us to rewrite Eq.~\ref{def} as:
\begin{equation}\label{eq:grad_mod}
\nabla =  \frac{1}{g}  \cdot \frac{{\cal R}}{\mu}    \cdot  \frac{dT}{dz}. 
\end{equation}
For simplicity, we define the coordinate system such that \( z \) increases toward deeper layers of the atmosphere, ensuring that \( \nabla \) and \( dT/dz \) have the same sign.

\begin{figure}\label{fig:app}
\includegraphics[width=\linewidth]{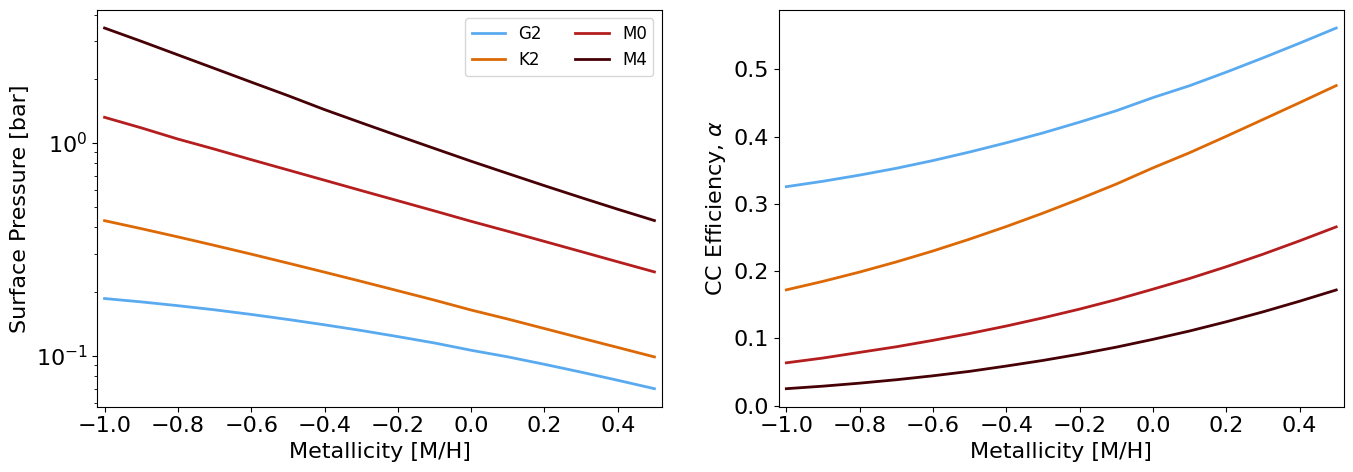}
\caption{Pressure at the stellar surface increases towards cooler and metal-poor stars (left panel).  Assuming that field strength at the surface of the flux tube does F depend on stellar parameters (see text) this leads to the decrease of the convective collapse efficiency for these stars (right panel).}
\end{figure}

Let us now use the Eddington approximation:
\begin{equation}\label{eq:Edd}
T^4 = 0.75 \cdot T_{\text{eff}}^4 \left( \tau_{\text{Ross}} + \frac{2}{3} \right),
\end{equation}
where $\tau_{\text{Ross}}$ is the Rosseland optical depth.
Differentiating it along the vertical direction yields:
\begin{equation}\label{eq:Edd_der}
4T^3 \cdot \frac{dT}{dz} = 0.75 \cdot T_{\text{eff}}^4 \cdot  \rho \, \kappa_{\text{Ross}},
\end{equation}
where $\kappa_{\text{Ross}}$ is a Rosseland opacity  {\it per unit of mass}  (this is why multiplication by density is needed).

By expressing \( {dT}{/dz} \) from Equation~\ref{eq:Edd_der}, relating \( T \) and \( T_{\text{eff}} \) using Equation~\ref{eq:Edd}, and substituting into Equation~\ref{eq:grad_mod} along with the ideal gas law, we obtain:
\begin{equation}
 \nabla = \frac{1}{4} \frac{P (\tau_{\text{Ross}})}{g} \cdot \kappa_{\text{Ross}} \cdot \frac{1}{\tau_{\text{Ross}}+2/3}
\end{equation}



Hydrostatic equilibrium implies that $P/g=m_c$ (with $m_c$ being the column mass). If Rosseland opacity does not change much over the atmosphere then at the optical surface ($\tau_{\text{Ross}}=1$):  $m_c \approx 1 / \kappa_{\text{Ross}}$ and $m_c \cdot \tau_{\text{Ross}} \approx 1$ so that the {\it gradient does not depend on stellar parameters}. 

Below the optical surface, the rising temperature leads to a rapid increase in opacity (for example, H$^-$ opacity scales approximately as $T^9$; see Chapter 9 of \citealt{Mihalas1978}). As a result, the Rosseland mean opacity becomes much larger than $1/m_c$, causing the temperature gradient $\nabla$ to increase sharply with depth and exceed the adiabatic value, thereby triggering the onset of convection. Convection acts to reduce $\nabla$, which asymptotically approaches the adiabatic value deeper in the convective zone. Overall, the maximum value of $\nabla$ occurs just below the stellar surface, where it not only initiates convection but also regulates the efficiency of convective collapse.

\section{Convective Collapse}\label{app:B}
To provide context for the expected dependence of convective collapse efficiency, $\alpha$, on stellar parameters, we begin with a brief overview of the convective collapse phenomenon and the formation of magnetic flux tubes. Convective collapse operates below the stellar surface, in layers where energy transport is dominated by convection and Equation~\ref{eq:grad} no longer applies. It is initiated by the advection of near-surface magnetic field into the downflow lanes by convective motions. Once the magnetic field accumulates in these downflows, it suppresses convective mixing and the exchange of heat between the downflowing gas and the surrounding, non-magnetic plasma. As a result, the temperature of the gas within the magnetic downflow evolves nearly adiabatically. In a superadiabatic environment of ambient plasma (see above), this means that the downflowing gas becomes relatively cooler and denser than its surroundings, leading to its negative buoyancy. The leads to the instability described by \cite{Spruit1979}: buoyancy forces accelerate the downflow and lead to the pressure deficit relative to their surroundings, that, in turn,  draws in more gas from above. The inflowing material brings additional magnetic flux, which becomes further concentrated by the horizontal contraction of the downflow region. This leads to field amplification and the formation of a strong, evacuated flux tube.

The convective collapse stops when the magnetic pressure in the flux tube becomes strong enough to balance the pressure deficit, and the inflow of gas into the flux tube ceases. This pressure drop is caused by a downflow and is regulated by the superadiabaticity of the surrounding non-magnetic plasma. In other words, a larger superadiabaticity below the surface leads to a more efficient convective collapse \citep[see also][]{Rajaguru2002}. The superadiabaticity decreases toward cooler stars, since less superadiabaticity is needed to transport their lower luminosities. Similarly, superadiabaticity decreases toward low-metallicity stars (that also have higher surface pressures), as reduced atmospheric opacities in these stars lead to a shallower rise of the temperature gradient $\nabla$ below the surface. While these illustrative arguments do not provide a precise explanation for why the increase in surface pressure toward cooler and metal-poor stars is almost fully compensated by the decrease in superadiabaticity and the efficiency of convective collapse, they do indicate that this compensation is not coincidental. Instead, it appears to be a self-regulating outcome of the near-surface stellar structure. This structure is shaped by the effective temperature (which sets the amount of energy that must be transported through the atmosphere), the metallicity (which determines the opacity and hence the efficiency of radiative transfer), and the surface gravity (which links the column mass to the surface pressure).

\acknowledgements
This work was supported by the ERC Synergy Grant REVEAL and the ERC Advanced Grant WINSUN under the European Union’s Horizon 2020 research
and innovation program (grant nos. 101118581 and 101097844). NK acknowledges support by German Aerospace Center (DLR) grants “PLATO Data Center”
50OO1501 and 50OP1902. YCU acknowledges financial support through STFC grants ST/S000372/1
and ST/W000989/1 AIS and TB acknowledge support by DFG grant SH1489/1. We acknowledge support by the Max Planck Computing and Data Facility. 


\end{document}